\documentstyle[graphicx]{mn}

\def\go{
\mathrel{\raise.3ex\hbox{$>$}\mkern-14mu\lower0.6ex\hbox{$\sim$}}
}
\def\lo{
\mathrel{\raise.3ex\hbox{$<$}\mkern-14mu\lower0.6ex\hbox{$\sim$}}
}
\def\simeq{
\mathrel{\raise.3ex\hbox{$\sim$}\mkern-14mu\lower0.4ex\hbox{$-$}}
}

\def\etal{{\it et al.\ }}

\def\etal{{\it et al.\ }}

\def\be{\begin{equation}}
\def\ee{\end{equation}}
\def\bea{\begin{eqnarray}}
\def\eea{\end{eqnarray}}

\def\etal{{\sl et al.\ }}

\def\hw2{{\hat W}^2}
\def\go{\mathrel{\raise.3ex\hbox{$>$}\mkern-14mu
             \lower0.6ex\hbox{$\sim$}}}

\def\etal{{\it et al.\ }}

\def\be{\begin{equation}}
\def\ee{\end{equation}}
\def\bea{\begin{eqnarray}}
\def\eea{\end{eqnarray}}

\def\etal{{\sl et al.\ }}

\def\lo{\mathrel{\raise.3ex\hbox{$<$}\mkern-14mu
             \lower0.6ex\hbox{$\sim$}}}
\def\ltorder{\mathrel{\raise.3ex\hbox{$<$}\mkern-14mu
             \lower0.6ex\hbox{$\sim$}}}
\def\gtorder{\mathrel{\raise.3ex\hbox{$>$}\mkern-14mu
             \lower0.6ex\hbox{$\sim$}}}

\def\eps2{{\epsilon^2}}

\begin{document}

\title[Detection of type-2 quasars in the radio galaxies B3~0731+438 and 3C~257]
{Detection of type-2 quasars in the radio galaxies B3~0731+438 and 3C~257}
\author[P.M. Derry \etal]{P.M. Derry$^{1}$, P.T. O'Brien$^{1}$, J.N. Reeves$^{1}$, M.J. Ward$^{1}$, M. Imanishi$^{2}$ \& S. Ueno$^{3}$ \\
$^{1}$ Department of Physics \& Astronomy, University of Leicester, University Road, Leicester, LE1 7RH, UK. \\
$^{2}$ National Astronomical Observatory of Japan, 2-21-1 Osawa, Mitaka, Tokyo 181-8588, Japan.\\
$^{3}$ Space Utilization Research Center, Tsukuba Space Center, National Space Development Agency of Japan, \\
2-1-1 Sengen, Tsukuba 305-8505, Japan.}

\date{Received ** *** 2002 / Accepted ** *** 2002}

\maketitle

\begin{abstract}
We present {\it XMM-Newton} observations and spectral fitting of two highly redshift, [OIII]-luminous, narrow-line radio galaxies, B3~0731+438 and 3C~257. Their X-ray continua are well fitted by a partial covering model with intrinsically unabsorbed and absorbed powerlaw components. The spectral models
indicate that both objects harbour highly obscured nuclei, with N$_{\rm H}$~$\approx$~0.5~$-$~2~$\times$~10$^{23}$~cm$^{-2}$. Correcting for this absorption we find large intrinsic luminosities in the range L$_{\rm X}$~$\approx$~0.2~$-$~1~$\times$~10$^{45}$~erg~s$^{-1}$. Thus, both sources are type-2 quasars. 

\end{abstract}


\section{Introduction}
\label{sec:intro}
An important issue in AGN research is whether heavily obscured powerful AGNs (type-2 quasars) are common. The unification of AGNs (Antonucci 1993), UV spectral shapes (Maiolino {\it et al.} 2001) and the modeling of AGNs infrared spectral energy distributions ({Pier \& Krolik 1993}) all suggest the presence of numerous type-2 quasars in the Universe. These obscured AGN are thought to outnumber unobscured objects by a factor of $\approx$~3:2 at redshifts out to z~$\approx$~3.5 (Willott {\it et al.} 2000; Fiore {\it et al.} 2001) and it is important to understand whether this ratio applies at higher redshifts and luminosities (Barcons {\it et al.} 1998; Almaini, Lawerence \& Boyle 1999; Stern {\it et al.} 2002). Obscured AGN are also believed to be responsible for a significant fraction of the hard X-ray background (Comastri {\it et al.} 1995). Almost all of the soft X-ray background ($<$~2~keV) has been resolved into discrete sources, the majority of which are broad line (BL) type-1 QSO's. At higher energies, typical unobscured QSO spectra are too steep to explain the XRB and account for only $\approx$~20~$\%$ of the sources (Hasinger {\it et al.} 1998). Obscured AGNs provide a natural explanation, since photoelectric absorption allows only the hard X-rays to penetrate. This implies the majority of the energy density generated by accretion in the Universe takes place in obscured AGN. These results have been confirmed by data from two deep {\it Chandra} observations (Brandt {\it et al.} 2001; Rosati {\it et al.} 2002) and recent analysis from an {\it XMM-Newton} observation of the Lockman Hole has also revealed similar results; Mainieri {\it et al.} (2002) found $\approx$~27~$\%$ of the Lockman Hole sample are Extremely Red Objects, the majority of which contain an obscured AGN. 

\begin{table*}
\begin{center}
\begin{tabular}{|c|l|c|l|c|l|c|}
\hline
Model &  & $\Gamma$ & N$_{\rm H}$ & $\chi$$^{2}$/d.o.f & f-test\\
& & & {\footnotesize (10$^{22}$ cm$^{-2}$)} & & $\%$ \\
\hline
& B3~0731+438 & & &\\
1 	&  po $\times$ wabs 		      & 1.42$^{+0.15}_{-0.15}$	& 0.06	& 29.15/32	& ----- \\
2 	& po $\times$ wabs $\times$ zwabs    & 1.83$^{+0.54}_{-0.10}$ & 3.3 $^{+2.5}_{-2.6}$ & 26.11/31	& 93.30\\
3 	& wabs(C*po $+$ zwabs $\times$ po)      & 2.61$^{+0.44}_{-0.37}$ & 18.8$^{+6.9}_{-5.5}$ & 16.89/30	& 99.97\\
\hline
& 3C~257 & & &\\
1 	& po $\times$ wabs 		      & 1.03$^{+0.11}_{-0.13}$ & 0.04 & 7.15/12	& ----- \\
2 	&  po $\times$ wabs $\times$ zwabs     & 1.46$^{+0.20}_{-0.25}$ & 5.0 $^{+2.9}_{-2.5}$ & 4.04/11	& 98.60\\
3 	& wabs(C*po $+$ zwabs $\times$ po)     & 1.42$^{+0.30}_{-0.21}$ & 5.5 $^{+6.4}_{-2.7}$	& 3.96/11	& 98.60\\
\hline
\end{tabular}
\end{center}
\caption{ The Models used to fit the 0.3~--~10~keV {\it XMM-Newton} data. For B3~0731+438 the scattering efficiency constant, C in Model 3, was found to be 11~$\%$~$\pm$ 6~$\%$ and for 3C~257 it was fixed at 10~$\%$. Also included in the table are the f-test results from comparing fits 2 and 3 with fit 1. One sigma errors are quoted.}
\end{table*}

Recently a small number of type-2 quasars candidates have been identified, but these results are usually based on limited data (Fabian {\it et al.} 2000; Willott {\it et al.} 2001; Fabian {\it et al.} 2001). The best published type-2 quasars, all obtained
using the Chandra X-ray Observatory, are IRAS~09104+4109 at z~$=$~0.44 (Iwasawa {\it et al.} 2001), originally observed with BeppoSAX (Franceschini {\it et al.} 2000), CDF-S~202 at z~$=$~3.7 (Norman {\it et al.} 2001), A18 at z~$=$~1.467 (Crawford {\it et al.}
2001), CX052, z~$=$~3.288 (Stern {\it et al.} 2002) and B2~0902+343, z~$=$~3.395 (Fabian, Crawford \& Iwasawa 2002). Correcting for absorption, these objects have X-ray luminosities of the order 10$^{45}$~erg~s$^{-1}$ and intrinsic absorption N$_{\rm H}$~$\approx$~10$^{24}$~cm$^{-2}$.

To identify likely type-2 quasars which are bright enough to provide reasonable X-ray spectra, we have observed two Narrow line radio galaxies (NLRGs) with {\it XMM-Newton}. The correlation between the mass of the central SMBHs and
the galaxy spheroidal component (Magorrian {\it et al.} 1998) indicates that SMBHs are most likely to reside in massive spheroids. Radio galaxies almost always host (forming) massive spheroids (van Bruegel {\it et al.} 1998), and NLRGs lack
broad components to their optical emission lines, so these galaxies are promising candidates for hosting obscured SMBHs (Barthel 1989). To identify such objects several parameters need to be measured; in particular the absorbing
column of the obscuring material and the intrinsic X-ray luminosity. In the context of this paper Type-2 quasars are defined to have an intrinsic, hard X-ray (2~--~10~keV) luminosity~$>$~10$^{44}$~erg~s$^{-1}$ (we assume cosmological values H$_{\rm o}$~$=$~75~km~s$^{-1}$~Mpc$^{-1}$ and q$_{\rm o}$~$=$~0.5 throughout); Less luminous objects are called Seyfert galaxies. The X-ray absorbing (gas and dust) column density of Seyfert 2s has N$_{\rm H}$~$\approx$~10$^{22}$~-~10$^{25}$~cm$^{-2}$ (Risaliti, Maiolino \& Salvati 1999), and type-2 quasars are expected to have similar values.  

We present {\it XMM-Newton} observations and spectral fitting of two powerful FRII radio galaxies, B3~0731+438 and 3C~257. Both sources are high redshift NLRGs with predicted X- ray luminosities well within the range of that expected
for type-2 quasars. 3C~257 is the highest redshift (z~$=$~2.474) radio galaxy observed in the 3C catalog with a fairly smooth K band emission and morphology resembling an elliptical galaxy (van Breugel 1998). B3~0731+438 (z~$=$~2.429) is a
typical FRII radio galaxy with double-lobed, radio emitting hot spots and a central core (Carilli {\it et al.} 1997; Motohara {\it et al.} 2000). 

This paper is organized as follows; in Section 2 we discuss sample selection and the use of optical emission line luminosities to determine the predicted X-ray luminosity of both sources. This is followed in Sections 3 and 4 by the {\it XMM-Newton} observations and data analysis. In Section 5 we discuss the spectral fitting and in the final section we conclude our findings.

\section{Target Selection}
\label{sec:Selection}
There is an observed correlation between L$_{\rm [OIII]}$ ($\lambda$5007) and the X-ray luminosity L$_{\rm X}$~(2~--~10~keV) in radio loud AGNs (Mulchaey {\it et al.} 1994; Sambruna, Eracleous \& Mushotzky 1999). This relation indicates NLRGs with L$_{\rm [OIII]}$ $>$ 5~$\times$~10$^{43}$~erg~s$^{-1}$ will have an X-ray luminosity L$_{\rm X}$~(2~--~10~keV)~$>$~5~$\times$~10$^{44}$~erg~s$^{-1}$ at the 90~$\%$ probability level. The relationship was used to select target NLRGs which are likely to host luminous AGNs and the two most luminous in {\rm [OIII]} were chosen for observation. From the [OIII] luminosities (Evans 1998) the predicted X-ray luminosities for B3~0731+438 (L$_{\rm [OIII]}$~$=$~12.4~$\times$~10$^{43}$~erg~s$^{-1}$) and 3C~257 (L$_{\rm [OIII]}$~$=$~14.0~$\times$~10$^{43}$~erg~s$^{-1}$) were estimated to be greater than 12 and 14~$\times$~10$^{44}$~erg~s$^{-1}$ respectively at the 90~$\%$ probability level. The width of the H$\alpha$ line in the K-band spectra was determined for both sources (Evans 1998; Jackson and Rawlings 1997); B3~0731+438 shows no broad H$\alpha$ component (FWHM$_{\rm H\alpha +[NII]}~$800 km s$^{-1}$) and 3C~257 has a somewhat broader line (FWHM$_{\rm H\alpha +[NII]}~$1800 km s$^{-1}$). Thus, for both objects their spectra indicate prominent emission from a narrow-line region while any nuclear emission is highly obscured.

For B3~0731+438 the diagnostic emission-line ratios are consistent with the presence of a Seyfert-2 nucleus (Evans 1998). A narrow-band optical image shows a large Ly$\alpha$ emission nebula (McCarthy 1991), suggestive of a hidden nucleus ionising the extended gas. This possibility is supported by a narrow-band 2.25~$\mu$m image obtained using the Subaru telescope showing biconical clouds extending out to 37 kpc, radiating H$\alpha$+[NII] emission lines with a total H$\alpha$+[NII] flux of 3.5~$\times$~10$^{-15}$~erg~s$^{-1}$~cm$^{-2}$ (Motohara {\it et al.} 2000), corresponding to a luminosity of 6.7~$\times$~10$^{43}$~erg~s$^{-1}$, (correcting to a cosmological scale of H$_{\rm o}$~$=$~75~km~s$^{-1}$~Mpc$^{-1}$ and q$_{\rm o}$~$=$~0.5). The H$\alpha$+[NII] ionization cone, aligned with the axis of the radio hot spots, implies the existence of a hidden, more luminous central source not observed in the optical, consistent with Keck spectropolarimetry data which suggest the nucleus is extinguished and we are only seeing a young stellar population in direct optical light (Vernet {\it et al.} 2001). The combination of these data suggest B3 0731+438 is a powerful, radio-loud AGN with a large intrinsic absorption. Little is known about 3C 257.

\section{Observations and Data Reduction}
\label{sec:Data}
B3~0731+438 and 3C~257 were observed during revolutions 343 (22nd October 2001) and 370 (15th December 2001) of {\it XMM-Newton} respectively. The observations of the EPIC MOS and PN detectors for both sources were in full window mode using a medium filter. The data were screened with the XMM Science Analysis Software (SAS) and pipeline processed. X-ray events corresponding to patterns 0~-~12 for the two MOS cameras and patterns 0~-~4, corresponding to single and double events for the PN were selected.

We then extracted the spectra and lightcurves for both the source and background in each observation, separately for each EPIC detector. Circular source regions of 1 arcmin diameter for B3~0731+438 and 0.6 arcmin for 3C~257 were selected. The background regions were placed at positions close to the sources and any significant flaring in the source and background regions were then removed using time selections in XMMSELECT. We then coadded the MOS1 and MOS2 data into a single spectral file to maximize the signal-to-noise ratio and grouped the data into a minimum of 20 counts per bin. 

The total effective observing time for B3~0731+438 is 22.75 ksec for PN and 27.40 ksec for both MOS1 and MOS2 with a source count rate of 0.01~cts/s for PN and 0.0033~cts/s for the coadded MOS data. The effective observing time for 3C~257 is 25.73 ksec for PN and 30.15 ksec for both MOS1 and MOS2, with a source count rate 0.0069~cts/s for PN and 0.0028~cts/s for the coadded MOS data. Finally, the background subtracted spectra were fitted using XSPEC version 11.1.0ae with the latest response matrices. 

\section{Spectral Analysis}
\label{sec:Analysis}

The spectral analysis was carried out in the conventional way. The data were compared with a range of parametric models until the best fit (best reduced $\chi$$^{2}$) was obtained. Seperate PN and MOS fits were carried out as a standard check for the fit and it was found that the data produced similar results to that of the coadded PN and MOS data. Hereafter, we present fits only to the coadded data with one sigma errors quoted throughout.

\subsection{B3~0731+438}

Initially we fitted the coadded PN and MOS spectra (0.3~--~10~keV) with a single powerlaw and neutral absorption (Galactic line-of-sight absorption) model; Model 1 in Table 1. Table 1 also includes values for the powerlaw photon index ($\Gamma$), the associated intrinsic column density (N$_{\rm H}$), where appropriate, and values for $\chi$$^{2}$ and number of degrees of freedom in the fit. Using a galactic column density N$_{\rm H}$ =$ $ 6.2~$\times$~10$^{20}$~cm$^{-2}$, Model 1 gives $\Gamma$~$=$~1.4~$\pm$~0.15~($\chi$$^{2}$/d.o.f~$=$~29.15/32), suggesting some intrinsic absorption. The unabsorbed flux and luminosity (2~--~10~keV) were found from the fit to be 3.2~$\times$~10$^{-14}$~erg~s$^{-1}$~cm$^{-2}$ and 2.8~$\times$~10$^{44}$~erg~s$^{-1}$ respectively.

The fit is improved (93~$\%$ significance) when intrinsic absorption is included in the spectral model; Model 2. This produces a steeper underlying powerlaw, $\Gamma$~$\approx$~1.8 and N$_{\rm H}$~$\approx$~3.3~$\times$~10$^{22}$~cm$^{-2}$ ($\chi$$^{2}$/d.o.f~$=$~26.11/31). This result suggests the source has a possible intrinisically scattered component, typical of a type-2 AGN. Therefore a partial covering model was fitted; Model 3, consisting of intrinsically absorbed and unabsorbed (scattered) powerlaw components. Model 3 provides an improved spectral fit for B3~0731+438 at greater than 99.9~$\%$ significance, with $\Gamma$~$=$~2.6~$\pm$ 0.4 and N$_{\rm H}$~$=$~1.9~$\times$~10$^{23}$~cm$^{-2}$~$\pm$ 0.7 ($\chi$$^{2}$/d.o.f~$=$~16.89/30). The unabsorbed X-ray luminosity (2~--~10~keV) is 9.1~$\times$~10$^{44}$~erg~s$^{-1}$. Figure 1 displays the best fit (Model 3) for B3~0731+438; the partial covering model. Figure 2 illustrates a contour plot of Model 3, showing the column density and X-ray luminosity distribution. The scattering efficiency was found to be 11~$\%$~$\pm$ 6~$\%$. Constraining the scattering efficiency to 10~$\%$ we find the results are consistent with the unconstrained fit. These results are consistent with the low scattering efficiencies for Seyfert 2s (Lumsden {\it et al.} 2001). 

The spectra of B3~0731+438 may be consistent with a reflection dominated component, but fitting the data with the reflection model PEXRAV (Magdziarz \& Zdziarski 1995) requires a very steep photon index of $\Gamma$~$\approx$~3.3~$\pm$~0.2 and no direct view of the incident continuum. If we assume the reflected component is $\approx$~5 $\%$, the implied upper limit on the intrinsic X-ray luminosity (2~--~10~keV), is 5.6~$\times$~10$^{45}$~erg~s$^{-1}$ if the planar reflection subtends 2$\pi$ sr at the illuminating source. Fitting the data with a broad (1~keV) FeK$\alpha$ emission line, we find a 90 $\%$ confidence upper limit on the intensity to be 2.7~$\times$~10$^{-6}$ photon s$^{-1}$~cm$^{-2}$, corresponding to a rest-frame equivalent width of 263 eV. This equivalent width is low, particularly for a type-2 object (Nandra {\it et al.} 1997; Turner {\it et al.} 1997).  

\begin{figure}
\begin{center}
\includegraphics[width=5cm,angle=-90]{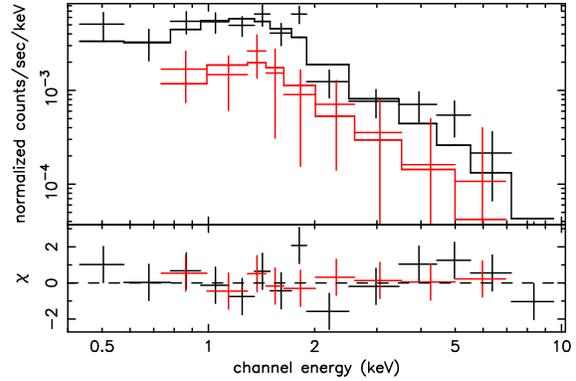}
\caption{The 0.3~--~10~keV EPIC coadded MOS (MOS1 + MOS2) and PN spectrum of B3~0731+438 fitted with a partial covering model (Model 3 in Table 1). The lower plot shows the ratio of the data to the fitted model. For clarity, the data have been rebinned to 30 counts per bin.}
\label{a}
\end{center}
\end{figure}

\begin{figure}
\begin{center}
\includegraphics[width=5cm,angle=-90]{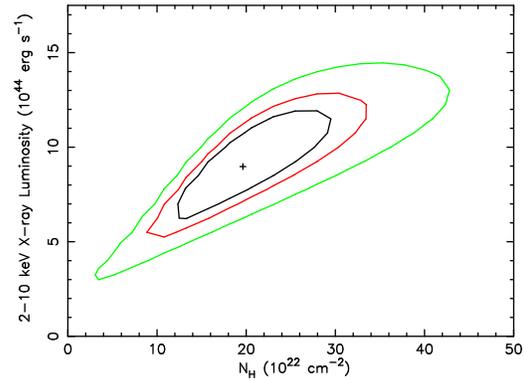}
\caption{A contour plot of column density against X-ray luminosity for B£~0731+438 from Model 3. The confidence intervals shown are for 68, 90 and 99 $\%$.}
\label{a}
\end{center}
\end{figure}

\subsection{3C~257}
The analysis of the PN data for 3C~257 was complicated by a residual spectral feature at $\approx$~1.5~keV due to incomplete subtraction of an internal detector background aluminium K-alpha line (Lumb {\it et al.} 2002). This line was removed from the spectral fitting by ignoring energies in the range 1.4~--~1.6~keV for the 3C 257 PN data during the analysis. 

The models used for B3~0731+438 were again used for 3C~257. Model 1, the single powerlaw and neutral absorption model, using a Galactic value of N$_{\rm H}$~$=$~4.4~$\times$~10$^{20}$~cm$^{-2}$, produced a very hard photon index $\Gamma$~$=$~1.0~$\pm$~0.1 ($\chi$$^{2}$/d.o.f~$=$~7.15/12). The flux and unabsorbed luminosity (2~--~10~keV) obtained from the fit were 2.6~$\times$~10$^{-14}$~erg~s$^{-1}$~cm$^{-2}$ and 1.5~$\times$~10$^{44}$~erg~s$^{-1}$ respectively. As in the case of B3~0731+438 the fit for 3C~257 is improved, with a 90 $\%$ significance, when intrinsic absorption is included in the spectral model; Model 2. This again produced a steeper underlying powerlaw with $\Gamma$~$\approx$~1.5 and N$_{\rm H}$~$\approx$~5.0~$\times$~10$^{22}$~cm$^{-2}$ ($\chi$$^{2}$/d.o.f~$=$~4.04/11).

We then used the partial covering model, Model 3, but it is not possible to distinguish between this model and Model 2. Since it was difficult to constrain the fit, we fixed the scattering efficiency to 10 $\%$, consistent with that of B3~0731+438. We found  $\Gamma$~$\approx$~1.4 and N$_{\rm H}$~$\approx$~5.5~$\times$~10$^{22}$~cm$^{-2}$ with a corrected, unabsorbed X-ray luminosity (2~--~10~keV)~$=$~2.6~$\times$~10$^{44}$~erg~s$^{-1}$. Figure 3 displays the spectrum of source 3C~257, fitted using  Model 3. Both models 2 and 3 suggest there is significant intrinsic absorption in 3C~257.

Fitting the data with the reflection model PEXRAV requires a steep photon index  $\Gamma$~$\approx$~2.7 with no direct view of the incident continuum. Making the same assumptions as for B3~0731+438 , the implied upper limit on the intrinsic X-ray luminosity (2~--~10~keV), is  3.0~$\times$~10$^{45}$~erg~s$^{-1}$.

\begin{figure}
\begin{center}
\includegraphics[width=5cm,angle=-90]{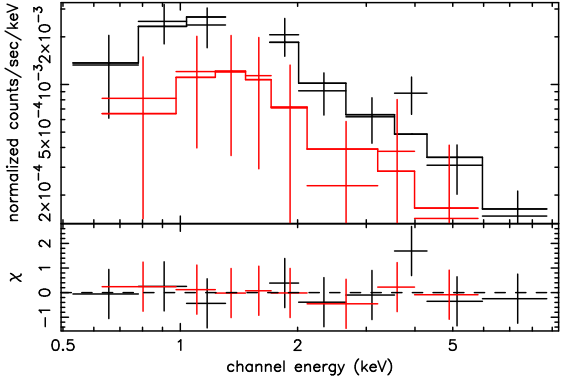}
\caption{{The 0.3~--~10~keV EPIC coadded MOS and PN spectrum of 3C~257 fitted with a partial covering model (Model 3, Table 1) binned to 20 counts per bin. The fit excludes the region from 1.4~--~1.6~keV (see text).}}
\label{c}
\end{center}
\end{figure}

\section{Discussion}
\label{sec:Disc}

The best-fit partial-scattering model for B3~0731+438 gives an intrinsic absorbing column of N$_{\rm H}$~$\approx$~1.9~$\times$~10$^{23}$~cm$^{-2}$ which is comparable to the column density suggested by Risaliti, Maiolino \& Salvati, (1999) for type-2 Seyferts of N$_{\rm H}$ $>$~10$^{23}$~cm$^{-2}$. This model also gave a photon index of $\Gamma$~$\approx$~2.6. This value is steep, but is comparable to the typical quasar value of $\Gamma$~$=$~1.9, at 90 $\%$ significance. Some luminous quasars have steep photon indices (e.g. $\Gamma$~$\approx$~2 for the quasar PDS 456 (Reeves {\it et al.} 1999)). For 3C~257 it was difficult to constrain the fit as a result of the limited number of photons. However, the spectral fits are consistent with this source also having a large X-ray luminosity and intrinsic absorption.

Both sources appear less luminous than predicted based on their [OIII] optical fluxes, but are still within the quasar range. The optical to X-ray spectral index $\alpha$$_{\rm ox}$ was determined for B3~0731+438 using IR spectroscopy from Eales \& Rawlings (1993) and Evans (1998). Assuming an optical extinction factor of A$_{\rm v}$~$=$~1.9 (Motohara {\it et al.} 2000) and using an absorption corrected X-ray flux, we calculate\\
 $\alpha$$_{\rm ox}$~$=$~-~1.87~$\pm$~0.2, consistent with sources studied by Bechtold {\it et al.} (2002).

For B3~0731+438 we can compare the ionizing photon number derived by Motohara {\it et al.} (2000) with the observed X-ray continuum flux. Integrating a powerlaw of photon index 2.5, consistent with that used by  Motohara {\it et al.} (2000), between 13.67 eV and infinity, normalised to the unabsorbed X-ray continuum, we derive a value of 1.7~$\times$~10$^{56}$~photons~s$^{-1}$. Repeating this calculation with our derived value of $\alpha$$_{\rm ox}$, extrapolating to 13.67 eV using the SED of Zheng {\it et al.} (1997), we derive a value of 1.0~$\times$~10$^{57}$~photons~s$^{-1}$. These photon numbers are consistent with the result obtained by Motohara {\it et al.} (2000), of $>$ 2.1~$\times$~10$^{56}$~photons~s$^{-1}$ based on the H$\alpha$ luminosity, assuming case B and a unity covering factor in the H$\alpha$ cone (corrected to a cosmological scale of H$_{\rm o}$~$=$~75~km~s$^{-1}$~ Mpc$^{-1}$ and q$_{\rm o}$~$=$~0.5). We can also compare the X-ray and  H$\alpha$ luminosities assuming the relation that the intrinsic X-ray luminosity (2~--~10~keV) is approximately ten times that of the H$\alpha$ luminosity (Ward {\it et al.} 1988; Ho {\it et al.} 2001). Again correcting to a cosmological scale of H$_{\rm o}$~$=$~75~km~s$^{-1}$~ Mpc$^{-1}$ and q$_{\rm o}$~$=$~0.5,  Motohara {\it et al.} (2000) found a H$\alpha$ luminosity for B3~0731+438 of 6.7~$\times$~10$^{43}$~erg~s$^{-1}$, giving a predicted X-ray luminosity of 6.7~$\times$~10$^{44}$~erg~s$^{-1}$. This is in good agreement with our value obtained for B3~0731+438 of  9.1~$\times$~10$^{44}$~erg~s$^{-1}$. These comparisons suggest the X-ray emission is fairly isotropic.

\section{Conclusions}
\label{sec:Con}
In this paper we have presented {\it XMM-Newton} spectral analysis for the two NLRGs B3~0731+438 and 3C~257. The analysis of  B3~0731+438 from fitting the spectra with a partial covering model indicates a spectral index of $\Gamma$~$=$~2.6 and a large column density N$_{\rm H}$~$\approx$~2~$\times$~10$^{23}$~cm$^{-2}$. As a result of few counts, it was difficult to constrain the fit for 3C~257. However, the results suggest a large column density with a hard photon index, in agreement with an obscured source. Both objects have large absorption corrected X-ray luminosities (2~--~10~keV of 9~$\times$~10$^{44}$~erg~s$^{-1}$ for B3~0731+438 and 2.6~$\times$~10$^{44}$~erg~s$^{-1}$ for 3C~257. 

In summary, the analysis has shown that both objects have intrinsic column densities and large X-ray luminosities well within the type-2 quasar range. We therefore conclude B3~0731+438 and 3C~257 are examples of type-2 quasar and currently provide the best X-ray spectra obtained of such objects with {\it XMM-Newton}. 

\section{ACKNOWLEDGMENTS}
The work in this paper is based on observations with {\it XMM-Newton}, an ESA science mission, with instruments and contributions directly funded by ESA and NASA. The authors would like to thank the EPIC Consortium for all their work during the calibration phase, and the SOC and SSC teams for making the observation and analysis possible. This research has made use of data obtained from the High Energy Astrophysics Science Archive Research Center (HEASARC), provided by NASA's Goddard Space Flight Center. PMD would also like to acknowledge support from a PPARC studentship.

\end{document}